\documentclass[11pt,english]{paper}
\usepackage{lmodern}

\usepackage[T1]{fontenc}
\usepackage[latin9]{inputenc}
\usepackage[letterpaper]{geometry}
\usepackage{amsthm}
\usepackage{amsmath}
\usepackage{amssymb}
\usepackage{graphicx}
\usepackage{esint}

\makeatletter

\providecommand{\tabularnewline}{\\}
\newcommand{\lyxdot}{.}

\numberwithin{equation}{section}
\numberwithin{figure}{section}
\numberwithin{table}{section}
\newcommand{\lyxaddress}[1]{
\par {\raggedright #1
\vspace{1.4em}
\noindent\par}
}

\makeatother

\usepackage{babel}
\begin{document}

\title{Navier-Stokes solver using Green's functions II: spectral integration
of channel flow and plane Couette flow}

\date{$\:$}

\author{Divakar Viswanath }

\maketitle

\lyxaddress{Department of Mathematics, University of Michigan (divakar@umich.edu). }
\begin{abstract}
The Kleiser-Schumann algorithm has been widely used for the direct
numerical simulation of turbulence in rectangular geometries. At the
heart of the algorithm is the solution of linear systems which are
tridiagonal except for one row. This note shows how to solve the Kleiser-Schumann
problem using perfectly triangular matrices. An advantage is the ability
to use functions in the LAPACK library. The method is used to simulate
turbulence in channel flow at $Re=80,000$ (and $Re_{\tau}=2400$)
using $10^{9}$ grid points. An assessment of the length of time necessary
to eliminate transient effects in the initial state is included.
\end{abstract}

\section{Introduction}

The incompressible Navier-Stokes equation \texttt{$\partial{\bf u}/\partial t+({\bf u}.\nabla){\bf u}=-\nabla p+\triangle{\bf u}/Re$},
where ${\bf u}$ is the velocity field and the pressure $p$ a Lagrange
multiplier for enforcing the incompressibility constraint $\nabla.{\bf u}=0$,
is an adequate physical model for a very great variety of phenomena
pertaining to fluid flows. The velocity field ${\bf u}=(u,v,w)$ is
represented as
\begin{equation}
{\bf u}=\sum_{l=-L/2}^{L/2}\sum_{n=-N/2}^{N/2}\hat{{\bf u}}_{l,n}(y)\exp\left(\frac{ilx}{\Lambda_{x}}+\frac{inz}{\Lambda_{z}}\right)\label{eq:secn1-fourier-of-u}
\end{equation}
with $\hat{{\bf u}}_{l,n}=\left(\hat{u}_{l,n},\hat{v}_{l,n},\hat{w}_{l,n}\right)$.
If either $L$ or $N$ is even, the $\pm L/2$ and $\pm N/2$ terms
are collapsed into a single cosine (or set to zero for convenience).
The three components of $\hat{{\bf u}}_{l,n}$ are represented using
Chebyshev polynomials $T_{n}(y)$ as $c_{0}/2+\sum_{j=1}^{M-1}c_{j}T_{j}+c_{M}T_{M}/2$
(it is typical to set $c_{M}$ to zero for convenience). The number
of grid points in the $y$ direction is always denoted by $M+1$.
This type of representation of velocity fields was first employed
by Orszag \cite{Orszag1971} in 1971.

The first computer simulation of fully developed turbulence in wall
bounded flows was reported in a paper by Kim, Moin, and Moser \cite{KimMoinMoser1987}.
Another effective and widely used algorithm for channel and plane
Couette geometries is due to Kleiser and Schumann \cite{KleiserSchumann1980}.
In this note, we give a reformulation of the Kleiser-Schumann algorithm.
This reformulation uses perfectly triangular matrices instead of triangular
matrices with a dense row. 

One of the largest $Re_{\tau}$ attained appears to be due to Hoyas
and Jiménez \cite{HoyasJimenez2006,HoyasJimenez2008}, who have reported
simulations at $Re_{\tau}=2003$. Hoyas and Jiménez use compact finite
differences to discretize derivatives in the $y$ direction. Unlike
the Chebyshev grid which is quadratically clustered at the endpoints
$y=\pm1$, the Hoyas-Jiménez grid is clustered like the $1.5$-th
power at $y=\pm1$. We report a turbulence simulation at $Re_{\tau}=2400$.

\section{A boundary value solver for $\left(D^{2}-a^{2}\right)u=f+\frac{dg}{dy}$}

In this section, we describe a boundary value solver for the linear
equation 
\begin{equation}
\left(D^{2}-a^{2}\right)u(y)=f(y)+\frac{dg(y)}{dy}\quad u(1)=A,\: u(-1)=B.\label{eq:secn2-bvp}
\end{equation}

The boundary value problem (\ref{eq:secn2-bvp}) is integrated once
with respect to $y$ to get the following equation.
\begin{equation}
Du-a^{2}\int\int Du=\int f+g.\label{eq:secn2-bvp-integrated}
\end{equation}
The solution derivative $Du$ is expanded in a truncated Chebyshev
series as $\alpha_{0}T_{0}(y)/2+\sum_{j=1}^{M-1}\alpha_{j}T_{j}(y)+\alpha_{M}T_{M}(y)/2$.
We use $\mathcal{T}_{j}(u)$ to denote the coefficient $\alpha_{j}$
in the Chebyshev series expansion of $u(y)$. For convenience, $\mathcal{T}_{M}(u)=\alpha_{M}$
is assumed to be zero. 

The first step is to find the particular solution of (\ref{eq:secn2-bvp-integrated})
subject to the integral conditions $\mathcal{T}_{0}(u)=\mathcal{T}_{0}(Du)=0$.
Because $\mathcal{T}_{0}(u)=0$, there is no indeterminate constant
in the Chebyshev series of $u$ obtained by integrating $Du$:
\begin{equation}
u=\int Du=\sum_{j=1}^{M-1}T_{j}\left(\frac{\alpha_{j-1}}{2j}-\frac{\alpha_{j+1}}{2j}\right)\label{eq:secn-chebseries-Du2u}
\end{equation}
where $\alpha_{0}=\alpha_{M}=0$. This equation is integrated once
more, once again using $\int T_{n}=T_{n+1}/2(n+1)-T_{n-1}/2(n-1)$
for $n>1$, $\int T_{1}=T_{2}/4$, and $\int T_{0}=T_{1}$, to get
\begin{align*}
Du-a^{2}\int\int u & =C+T_{1}\left(\alpha_{1}+a^{2}\left(\frac{\alpha_{1}}{8}-\frac{\alpha_{3}}{8}\right)\right)\\
 & +\sum_{j=2}^{M-1}T_{j}\left(\alpha_{j}-a^{2}\left(\frac{\alpha_{j-2}}{4j(j-1)}-\frac{\alpha_{j}}{2\left(j^{2}-1\right)}+\frac{\alpha_{j+2}}{4j(j+1)}\right)\right)
\end{align*}
where $\alpha_{0}=\alpha_{M}=\alpha_{M+1}=0$. Here $C$ is an indeterminate
constant.

Similarly, if $\mathcal{T}_{j}(f)=f_{j}$ and $\mathcal{T}_{j}(g)=g_{j}$
in the truncated Chebyshev series of the functions $f$ and $g$,
the expansion of $\int f+g$ is given by
\[
\int f+g=C+\sum_{j=1}^{M-1}T_{j}\left(\frac{f_{j-1}}{2j}-\frac{f_{j+1}}{2j}+g_{j}\right)
\]
where $C$ is another indeterminate constant and it is assumed that
$f_{M}=g_{M}=0$. Equating the Chebyshev coefficients on either of
side of (\ref{eq:secn2-bvp-integrated}) for $j=1,\ldots,M-1$ we
have
\begin{align}
\left(1+\frac{a^{2}}{8}\right)\alpha_{1}-\left(\frac{a^{2}}{8}\right)\alpha_{3} & =\frac{f_{0}}{2}-\frac{f_{2}}{2}+g_{1}\quad\text{for \ensuremath{j=1}, and}\nonumber \\
-\left(\frac{a^{2}}{4j(j-1)}\right)\alpha_{j-2}+\left(1+\frac{a^{2}}{2(j^{2}-1)}\right)\alpha_{j}-\left(\frac{a^{2}}{4j(j+1)}\right) & \alpha_{j+2}=\frac{f_{j-1}}{2j}-\frac{f_{j+1}}{2j}+g_{j}\label{eq:secn2-eqn-for-alphas}
\end{align}
for $j=2,3,\ldots,M-1$. These $M-1$ equations decouple into two
tridiagonal systems of dimensions $\frac{M-2+M\bmod2}{2}$ and $\frac{M-M\bmod2}{2}$
for the even and odd modes, respectively. The equations are solved
for $\alpha_{1},\ldots,\alpha_{M-1}$ to find the Chebyshev series
of $Du$. The Chebyshev series of $u$ is found using that of $Du$
(\ref{eq:secn-chebseries-Du2u}). This particular solution $u$ satisfies
the integral conditions $\mathcal{T}_{0}(Du)=\mathcal{T}_{0}(u)=0$.

The particular solution found in this way is quite inaccurate when
$a$ is large, which is the typical situation when the Reynolds number
$Re$ is large. However, the boundary value problem (\ref{eq:secn2-bvp})
can still be solved accurately. To do so, the homogeneous solutions
of (\ref{eq:secn2-bvp}) must be found in a peculiar way. For reasons
explained in \cite{Viswanath2014}, this peculiar way of finding homogeneous
solutions leads to a cancellation of discretization error and an accurate
solution of the original boundary value problem.

One of the homogeneous solutions is taken to be of the form $u_{h_{1}}=\frac{1}{2}+u$
with $u$ satisfying $\mathcal{T}_{0}(Du)=\mathcal{T}_{0}(u)=0$.
Then $u$ satisfies $\left(D^{2}-a^{2}\right)u=a^{2}/2$. Thus $Du_{h_{1}}$
and $u_{h_{1}}$ may be found by solving (\ref{eq:secn2-eqn-for-alphas})
with $g\equiv0$ and $f_{0}=a^{2},\, f_{1}=\cdots=f_{M}=0$.

The other homogeneous solution is taken to be of the form $u_{h_{2}}=T_{1}/2+u$
with $\mathcal{T}_{0}(Du)=\mathcal{T}_{0}(u)=0$ as before. This time
$\left(D^{2}-a^{2}\right)u=a^{2}T_{1}/2$ and (\ref{eq:secn2-eqn-for-alphas})
must be solved using $f_{1}=a^{2}/2$ and $f_{0}=f_{2}=\cdots=f_{M}=0$.

Finding the particular solution $u$ and the homogeneous solutions
$u_{h_{1}},\, u_{h_{2}}$ involves solving a total of $6$ tridiagonal
systems each of dimension approximately $M/2$. In the linear combination
$u+c_{1}u_{h_{1}}+c_{2}u_{h_{2}}$ the coefficients $c_{1}$ and $c_{2}$
are calculated by setting the boundary values at $\pm1$ to $A$ and
$B$ as in (\ref{eq:secn2-bvp}). The solution derivative is given
by $Du+c_{1}Du_{h_{1}}+c_{2}Du_{h_{2}}$. Here the derivatives $Du,\, Du_{h_{1}},\, Du_{h_{2}}$
are from solving the tridiagonal systems and not from numerical differentiation.
The function $g(y)$ on the right hand side of (\ref{eq:secn2-bvp})
is not differentiated numerically either.

\section{The Kleiser-Schumann algorithm}

Here we use the boundary value solver of Section 2 to derive a version
of the Kleiser-Schumann algorithm \cite{KleiserSchumann1980}, which
does not use numerical differentiation.

In the Fourier decomposition of the velocity field (\ref{eq:secn1-fourier-of-u}),
denote the Fourier mode $\hat{{\bf u}}_{l,n}$ by $(u,v,w)$, after
dropping the subscripts $l,n$. The Fourier mode at the end of time
step $q$ is denoted by $\left(u^{q},v^{q},w^{q}\right)$. Similarly,
if ${\bf H}={\bf \omega}\times{\bf u}$ is the nonlinear term, denote
its $l,n$-th Fourier mode by $\left(H_{1},H_{2},H_{3}\right)$. Similarly,
$p$ denotes the $l,n$-th Fourier mode of pressure (more precisely,
pressure plus $|{\bf u}|^{2}/2$) and $p^{q}$ denotes that Fourier
mode at the end of time step $q$.

The Navier-Stokes equation for the $l,n$-th mode takes the following
form:
\begin{align*}
\frac{\partial u}{\partial t}+H_{1} & =-\frac{il}{\Lambda_{x}}p+\frac{1}{Re}\left(D^{2}-\alpha^{2}\right)u\\
\frac{\partial v}{\partial t}+H_{2} & =-\frac{\partial p}{\partial y}+\frac{1}{Re}\left(D^{2}-\alpha^{2}\right)v\\
\frac{\partial w}{\partial t}+H_{3} & =-\frac{in}{\Lambda_{z}}p+\frac{1}{Re}\left(D^{2}-\alpha^{2}\right)w
\end{align*}
where $\alpha^{2}=l^{2}/\Lambda_{x}^{2}+n^{2}/\Lambda_{z}^{2}$. The
incompressibility condition is $(il/\Lambda_{x})u+\partial v/\partial y+(in/\Lambda_{z})w=0$.
Using the incompressibility condition, we get
\[
\left(D^{2}-\alpha^{2}\right)p=-\frac{il}{\Lambda_{x}}H_{1}-\frac{\partial H_{2}}{\partial y}-\frac{in}{\Lambda_{z}}H_{3}.
\]

Each of these equations is of the form $dX/dt=f(X)+\triangle X/Re$,
with $\triangle=\left(D^{2}-\alpha^{2}\right)$. The time discretizations
we consider are of the following form \cite{ViswanathTobasco2013}:
\begin{equation}
\frac{1}{\Delta t}\left(\gamma X^{q+1}+\sum_{j=0}^{s-1}a_{j}X^{q-j}\right)=\sum_{j=0}^{s-1}b_{j}f(X^{q-j})+\frac{1}{Re}\triangle X^{q+1}.\label{eq:secn3-time-disc}
\end{equation}
The method applies to other discretizations such as Runge-Kutta with
little change. Low storage Runge-Kutta schemes, which use only slightly
more memory than the Euler method, are more commonly used. The low-storage
scheme in \cite{SpalartMoserRogers1991} uses three sub-steps, is
$2$nd order in viscous terms and $3$rd order in convection terms.
The theoretical CFL limit is $\sqrt{3}$ \cite{SpalartMoserRogers1991}
or $\sqrt{3}/3\approx0.58$ per sub-step. In comparison, the $3$rd
order $3$ step scheme of the form above has $0.63$ as its theoretical
CFL limit. There appears to be room for a little progress here, which
is why we are opting for multistep methods.

If the equations for the $l,n$-th mode are time-discretized, we get
\begin{align}
\left(D^{2}-\beta^{2}\right)u & =Re\left(\tilde{H}_{1}+U\right)+\frac{il}{\Lambda_{x}}.p.Re\nonumber \\
\left(D^{2}-\beta^{2}\right)v & =Re\left(\tilde{H}_{2}+V\right)+\frac{dp}{dy}.Re\nonumber \\
\left(D^{2}-\beta^{2}\right)w & =Re\left(\tilde{H}_{3}+W\right)+\frac{in}{\Lambda_{z}}.p.Re,\label{eq:secn3-ks-eqns}
\end{align}
where $\beta^{2}=\alpha^{2}+\gamma Re/\Delta t$. Here $(u,v,w)$
is the $l,n$-th mode of the velocity field (compare (\ref{eq:secn1-fourier-of-u}))
at the end of time step $q+1$ and likewise with $p$. Here $\tilde{H}_{i}=\sum_{j=0}^{s-1}b_{j}H_{i}^{q-j}$
for $i=1,2,3$ and $U=\frac{1}{\Delta t}\sum_{j=0}^{s-1}a_{j}u^{q-j}$.
The coefficients $a_{j}$ and $b_{j}$ are from the time discretization
(\ref{eq:secn3-time-disc}). The quantities $V$ and $W$ are defined
in the same manner as $U$. The pressure Poisson equation takes the
form
\begin{equation}
\left(D^{2}-\alpha^{2}\right)p=-\frac{il}{\Lambda_{x}}\tilde{H}_{1}-\frac{\partial\tilde{H}_{2}}{\partial y}-\frac{in}{\Lambda_{z}}\tilde{H}_{3}.\label{eq:secn3-pressure-poisson}
\end{equation}

We will show how to solve (\ref{eq:secn3-pressure-poisson}) and (\ref{eq:secn3-ks-eqns})
without numerical differentiation in the $y$ direction. The quantities
\[
u,v,w,H_{1},H_{2},H_{3}
\]
are assumed to be available at the end of time steps $q,q-1,\ldots,q-s+1$.
They are used to compute $\tilde{H}_{i}$ and $U,V,W$. The first
step, described at greater length below, is to solve for $u,du/dy,v,dv/dy,w,dw/dy$
at the end of time step $q+1$ using (\ref{eq:secn3-pressure-poisson})
and (\ref{eq:secn3-ks-eqns}). Once these quantities are available
the vorticity component $\hat{{\bf \omega}}_{l,n}$ may be formed
using arithmetic operations and ${\bf H}=\omega\times{\bf u}$ is
available for the next time step.

The equations (\ref{eq:secn3-pressure-poisson}) and (\ref{eq:secn3-ks-eqns})
are solved for $u,du/dy,v,dv/dy,w,dw/dy$ as follows. 
\begin{itemize}
\item Find a particular solution $p^{\ast}$ of the pressure Poisson equation
(\ref{eq:secn3-pressure-poisson}) using the boundary value solver
of Section 2. Note that $g=-\tilde{H}_{2}$ when that solver is applied
and $\tilde{H}_{2}$ is not differentiated numerically.
\item Find two solutions $p_{1}$ and $p_{2}$ of the homogeneous part of
(\ref{eq:secn3-pressure-poisson}), which is $\left(D^{2}-\alpha^{2}\right)p=0$
as described in Section 2. These two solutions can be precomputed
and reused at every time step.
\item Assume $p=p^{\ast}+c_{1}p_{1}+c_{2}p_{2}$ and solve the $v$ equation
of (\ref{eq:secn3-ks-eqns}) with the boundary conditions $v(\pm1)=0$
using the boundary value solver of Section 2. Since $c_{1}$ and $c_{2}$
are undetermined constants, we may take the solution to be $v=v^{\ast}+c_{1}v_{1}+c_{2}v_{2}$,
where all of $v^{\ast},v_{1},v_{2}$ are zero at the walls. Note that
$p^{\ast},p_{1},p_{2}$ are not differentiated numerically because
the boundary value solver which produces those quantities also produces
their derivatives. Alternatively, numerical differentiation can be
avoided by using $p*,p_{1},p_{2}$ multiplied by $Re$ in place of
$g$ in the boundary value solver of Section 2.1. As long as the time
step $\Delta t$ does not change, we may precompute and reuse $v_{1}$
and $v_{2}$. The indeterminate quantities $c_{1}$ and $c_{2}$ are
calculated using the zero divergence condition $(il/\Lambda_{x})u+\partial v/\partial y+(in/\Lambda_{z})w=0$.
More specifically, we must have $dv/dy=0$ at $y=\pm1$ since the
no-slip boundary requires $u=v=w=0$ at the walls for all but the
mean mode. Note that $v^{\ast},v_{1},v_{2}$ are not differentiated
numerically to enforce $dv/dy=0$ at the walls. The boundary value
solver that produces those quantities also produces their derivatives.
\item Once we have $p$, we may solve the $u$ and $w$ equations of (\ref{eq:secn3-ks-eqns})
to produce $u,du/dy$ as well as $w,dw/dy$. 
\end{itemize}
To complete the description of this method, we show how the mean modes
are handled. The mean Fourier components are denoted using an over-bar.
The equations for the mean modes at the end of time step $q+1$ are
as follows:
\begin{align*}
\left(D^{2}-\frac{\gamma Re}{\Delta t}\right)\bar{u} & =Re\left(P+\tilde{H}_{1}+U\right)\\
\frac{d\bar{p}}{dy} & =-\tilde{H}_{2}\\
\left(D^{2}-\frac{\gamma Re}{\Delta t}\right) & \bar{w}=Re\left(\tilde{H}_{3}+W\right)
\end{align*}
Here $\tilde{H}_{i}=\sum_{j=0}^{s-1}b_{j}\bar{H}_{i}^{q-j}$, $U=\frac{1}{\Delta t}\sum_{j=0}^{s-1}a_{j}\bar{u}^{q-j}$,
and likewise for $W$. The contribution of the pressure gradient is
$P$, which is set to zero for plane Couette flow. For channel flow
the pressure gradient may be fixed at $-2/Re$, but turbulence is
better sustained by fixing the mass flux $\frac{1}{2}\int_{-1}^{+1}\bar{u}\, dy$
at $2/3$. The mass flux is maintained by choosing 
\[
p_{g}=-\frac{1}{2}\int_{-1}^{+1}\bar{H}_{1}\, dy+\frac{1}{2Re}\frac{\partial\bar{u}}{\partial y}\Biggl|_{y=-1}^{y=1}.
\]
In either case the laminar solution is ${\bf u}=(0,1-y^{2},0)$. In
the equations above, $P=\sum_{j=0}^{s-1}b_{j}p_{g}^{q-j}$. The boundary
condition for $\bar{u}$ is $\bar{u}(\pm1)=0$ for channel flow and
$\bar{u}(\pm1)=\pm1$ for plane Couette flow. These three equations
for the mean mode are uncoupled. The second equation is solved for
$\bar{p}$. From the other two equations, $\bar{u},\bar{w}$ as well
as their derivatives are obtained using the boundary value solver
of Section 2.1. The mean component $\bar{v}$ is zero as a consequence
of incompressibility of the fluid and the no-slip boundary.

\section{Numerical results}

The computations described in this section were run on $10$ compute
nodes, each a $2.67$ GHZ Xeon 5650 with $12$ processor cores, connected
over QDR Infiniband network. The peak memory bandwidth (for reading
data) on a single node is more than $30$ GB/s. The peak bidirectional
network bandwidth on a single node approaches $5$ GB/s. Each node
has $48$ GB of memory. The largest computation carried out used $10^{9}$
grid points to reach $Re_{\tau}=2380$. 

\begin{table}
\begin{centering}
\begin{tabular}{|c|c|c|c|c|c|c|c|c|}
\hline 
$Re$ & $Re_{\tau}$ & $\Lambda_{x}$ & $\Lambda_{z}$ & $L/M/N$ & $dx^{+}/dy_{max}^{+}/dz^{+}$ & $T_{init}u_{\tau}/h$ & $Tu_{\tau}/h$ & CFL\tabularnewline
\hline 
\hline 
$16875$ & $601$ & $2$ & $1$ & $384/320/384$ & $9.8/5.9/4.9$ & $\geq30$ & $40.86$ & $0.25$\tabularnewline
\hline 
\end{tabular}
\par\end{centering}

\caption{Run parameters for validation. Grid spacings in frictional units are
$dx^{+}/dy_{max}^{+}/dz^{+}$. Time of integration for eliminating
transients is $T_{init}$ and the time of integration for gathering
statistics is $T$. CFL is the typical Courant-Friedrichs-Lewy number
of a time step. \label{tab:secn5.1-validation}}
\end{table}

Moser, Kim, and Mansour (MKM) \cite{MoserKimMansour1999} carried
out direct numerical simulations of channel flow at $Re_{\tau}=587$.
The run parameters of Table \ref{tab:secn5.1-validation} are the
same as that of MKM with minor differences. The Reynolds number $Re_{\tau}=601$
is slightly higher and the grid in the wall-normal direction is finer
using $M=320$ instead of $M=256$. For second order statistics, the
grid resolutions are required to be $dx^{+}\approx9$, $dy_{max}^{+}\approx7$,
and $dz^{+}\approx5$.%
\footnote{Thanks to a referee of part I of this sequence for this valuable information.%
} The initial run to eliminate transients and achieve a statistically
steady turbulent state satisfies $T_{init}\geq30\tau_{eddy}$. Thirty
eddy turnover times is quite a long integration and we may be sure
that transients are thoroughly eliminated.

\begin{figure}
\begin{centering}
\includegraphics[scale=0.3]{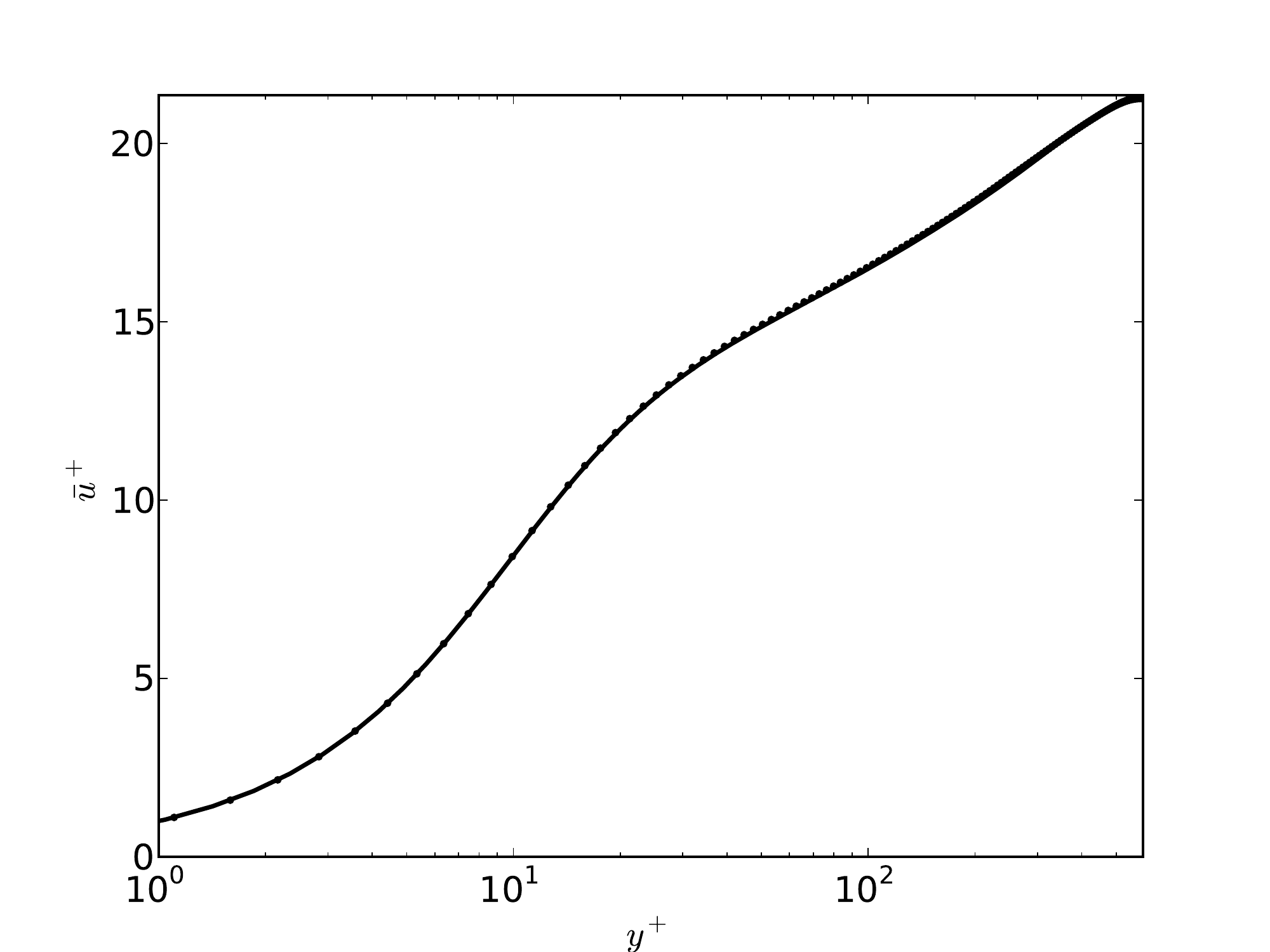}
\par\end{centering}

\begin{centering}
\includegraphics[scale=0.3]{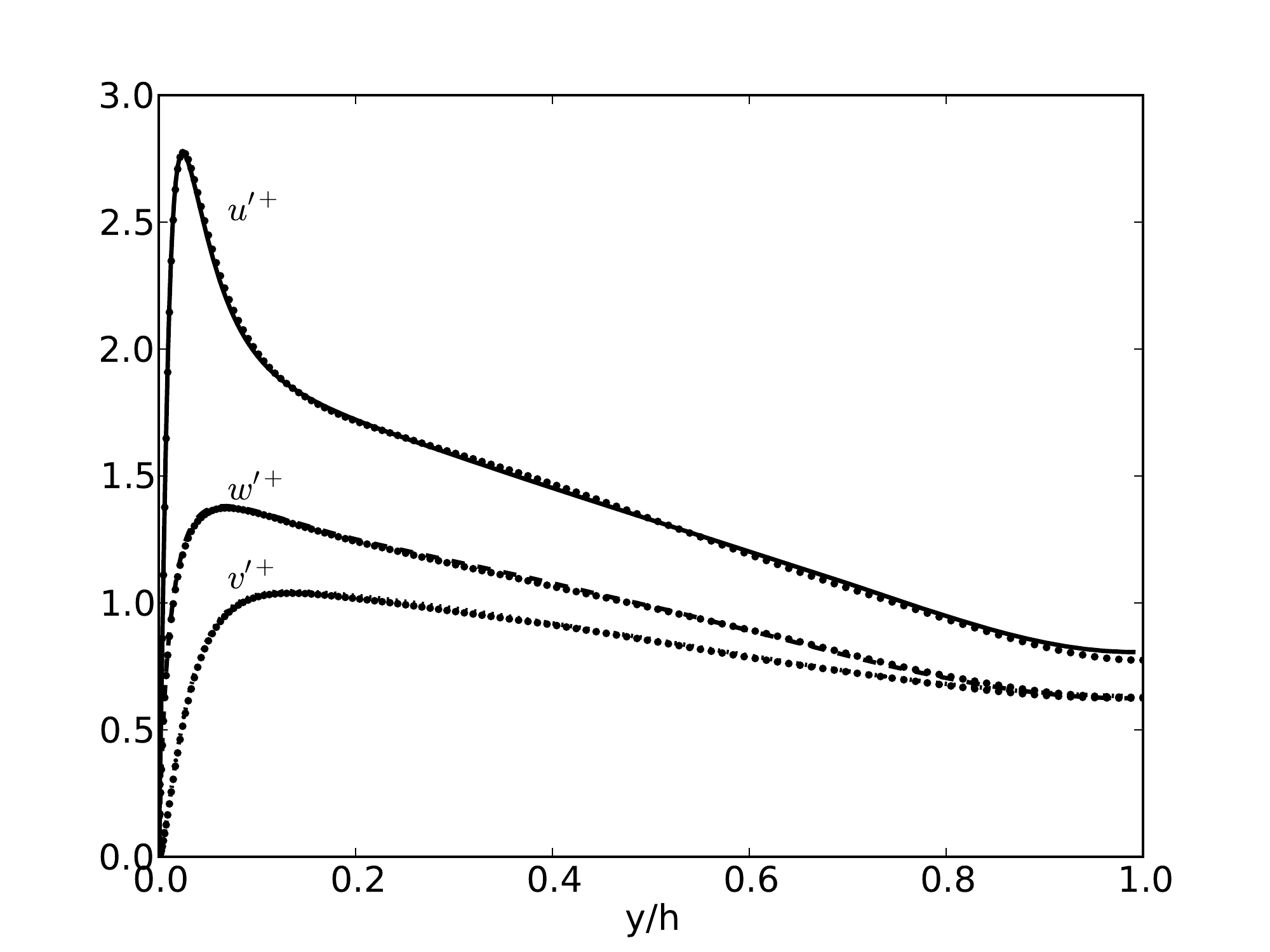}\includegraphics[scale=0.3]{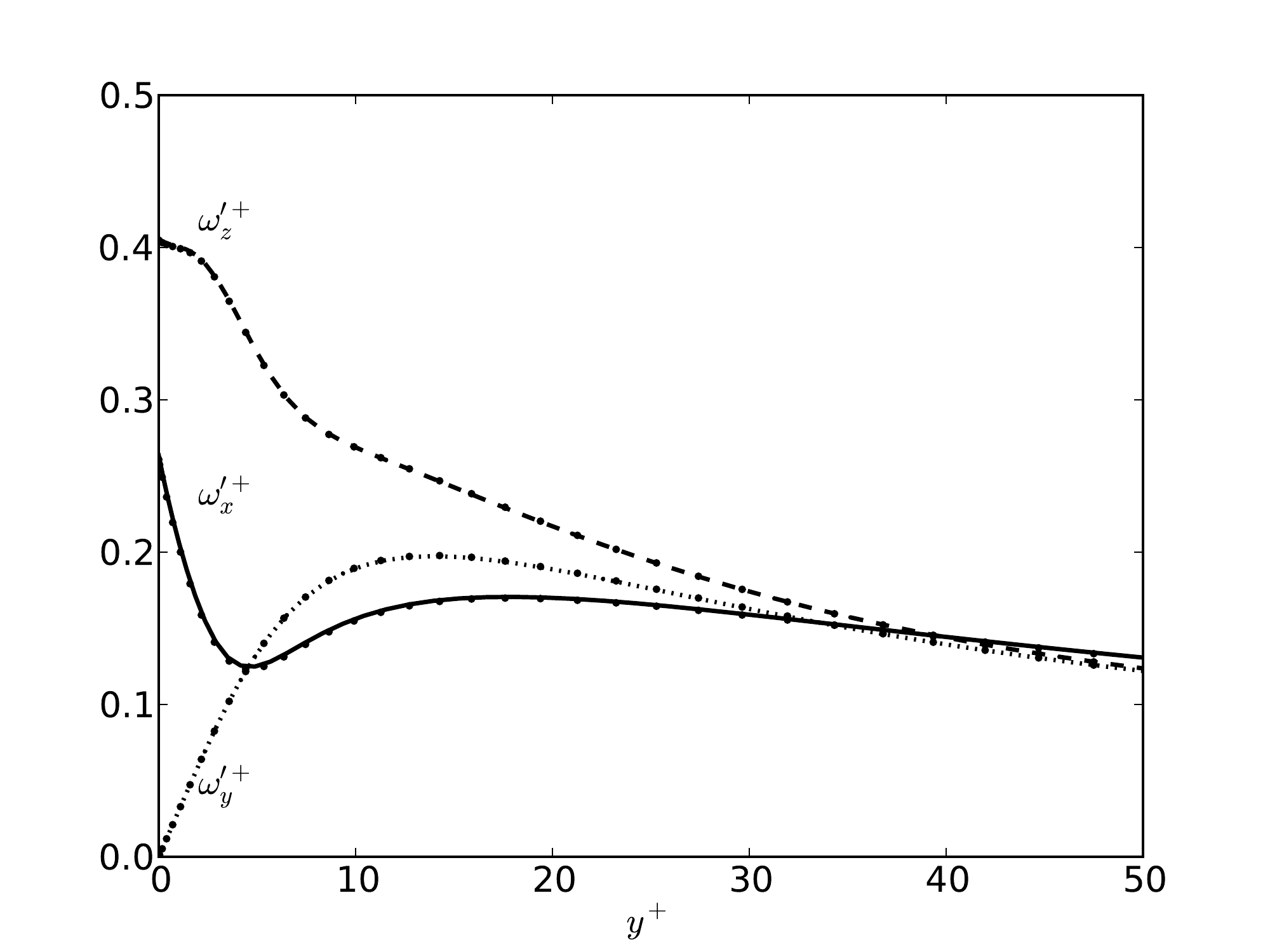}
\par\end{centering}

\centering{}\caption{Validation of run of Table \ref{tab:secn5.1-validation} against data
from \cite{MoserKimMansour1999}, which is dotted. The plots show
mean streamwise velocity, rms turbulence intensities, and rms vorticity
fluctuations as a function of the distance from the wall.\label{fig:secn5-validation-1}}
\end{figure}

Figure \ref{fig:secn5-validation-1} shows excellent agreement of
mean streamwise velocity, turbulence intensities, and vorticity fluctuations.
The slight discrepancy in streamwise turbulence intensity $u^{\prime+}$
visible near $y/h\approx0.4$ is most likely because the MKM run was
not as long as that of Table \ref{tab:secn5.1-validation}. A run
with $T>40\tau_{eddy}$ is affordable on even a small cluster purchased
in 2010, thanks to $15+$ years of exponential increase in computing
power, but would not have been affordable to MKM. The slight discrepancy
in streamwise turbulence intensity near $y/h\approx1$ is probably
because the $Re_{\tau}$ in Table \ref{tab:secn5.1-validation} is
slightly higher than that of MKM.

\begin{figure}
\begin{centering}
\includegraphics[scale=0.3]{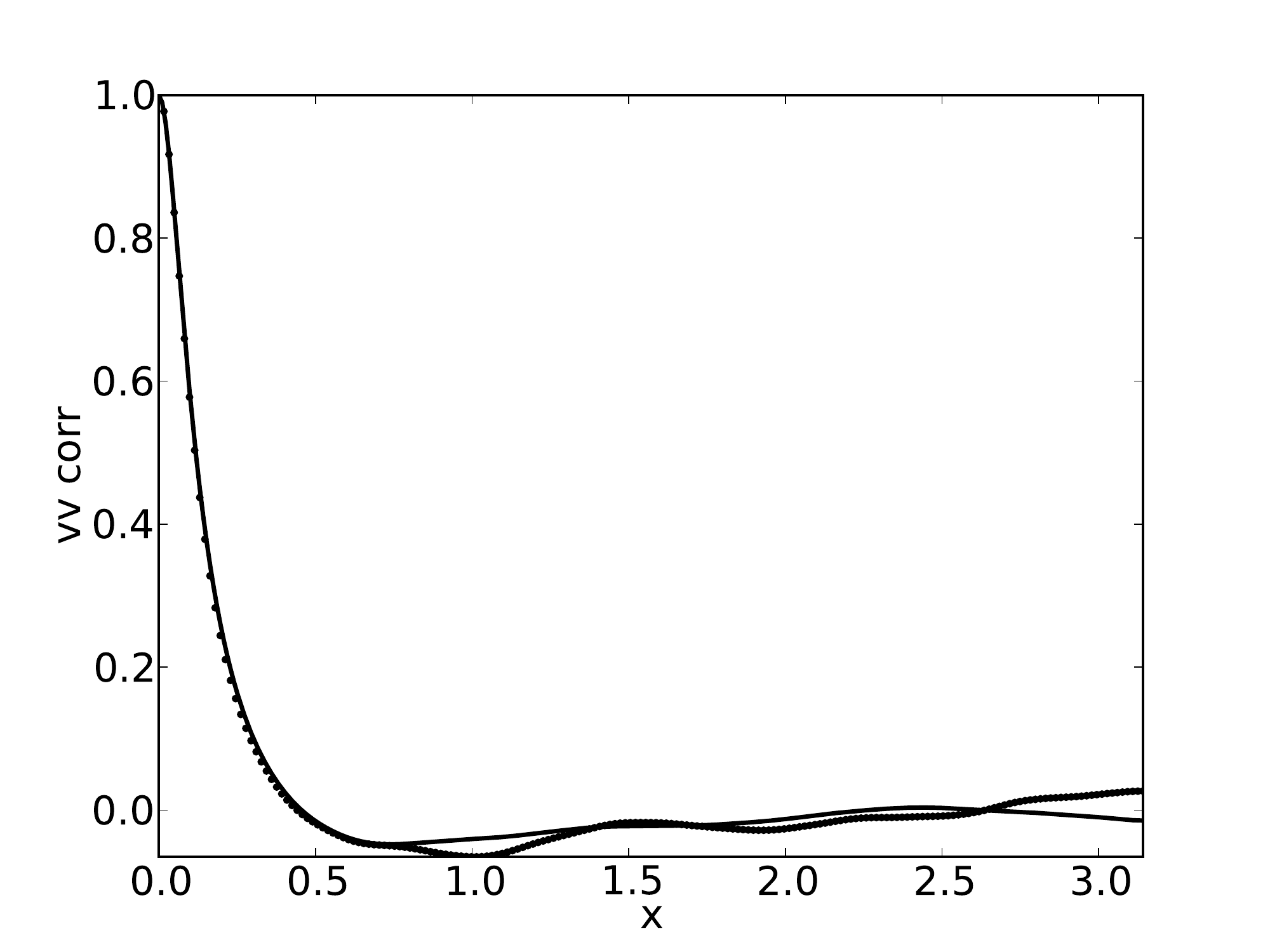}\includegraphics[scale=0.3]{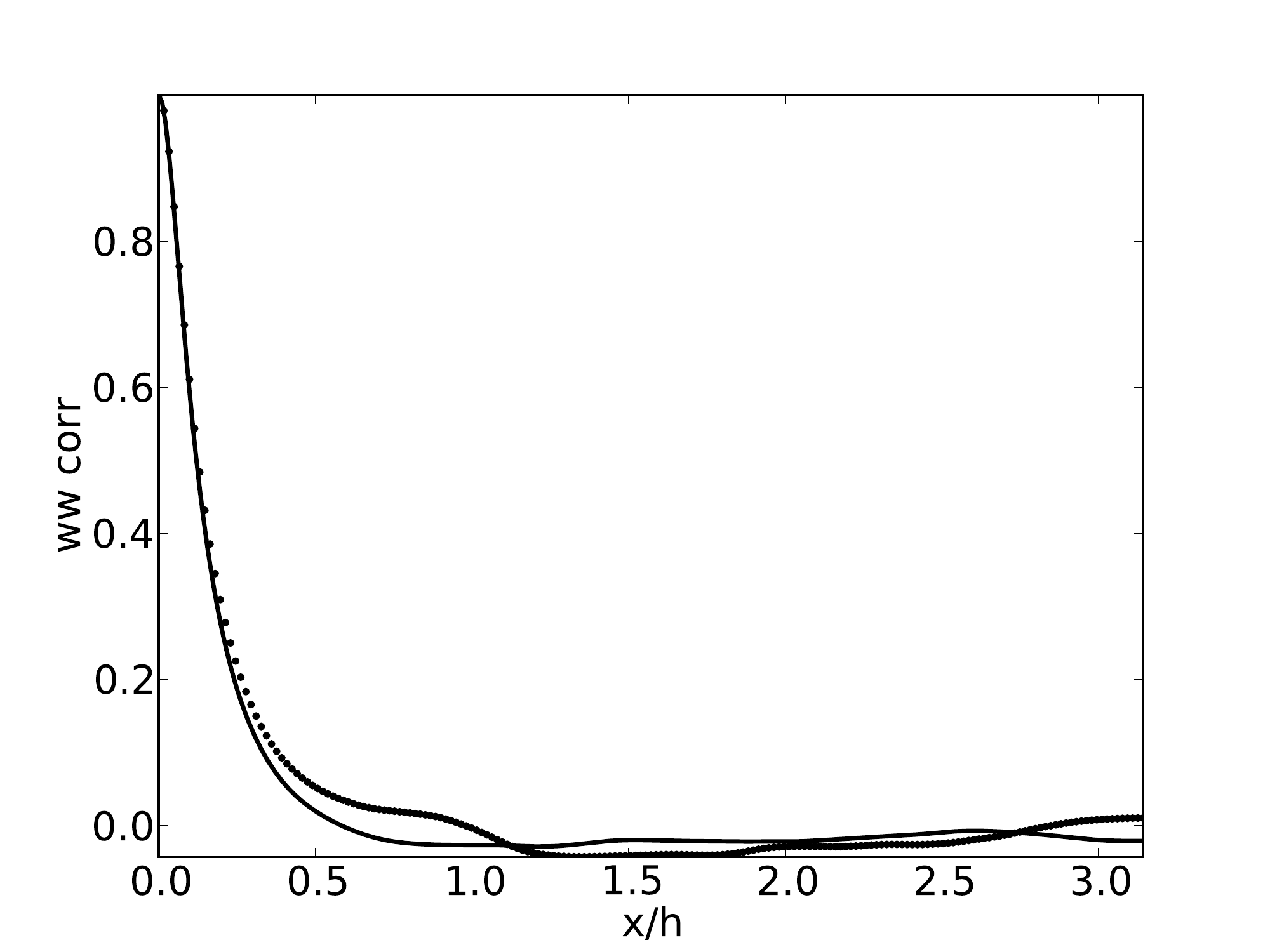}
\par\end{centering}

\begin{centering}
\includegraphics[scale=0.3]{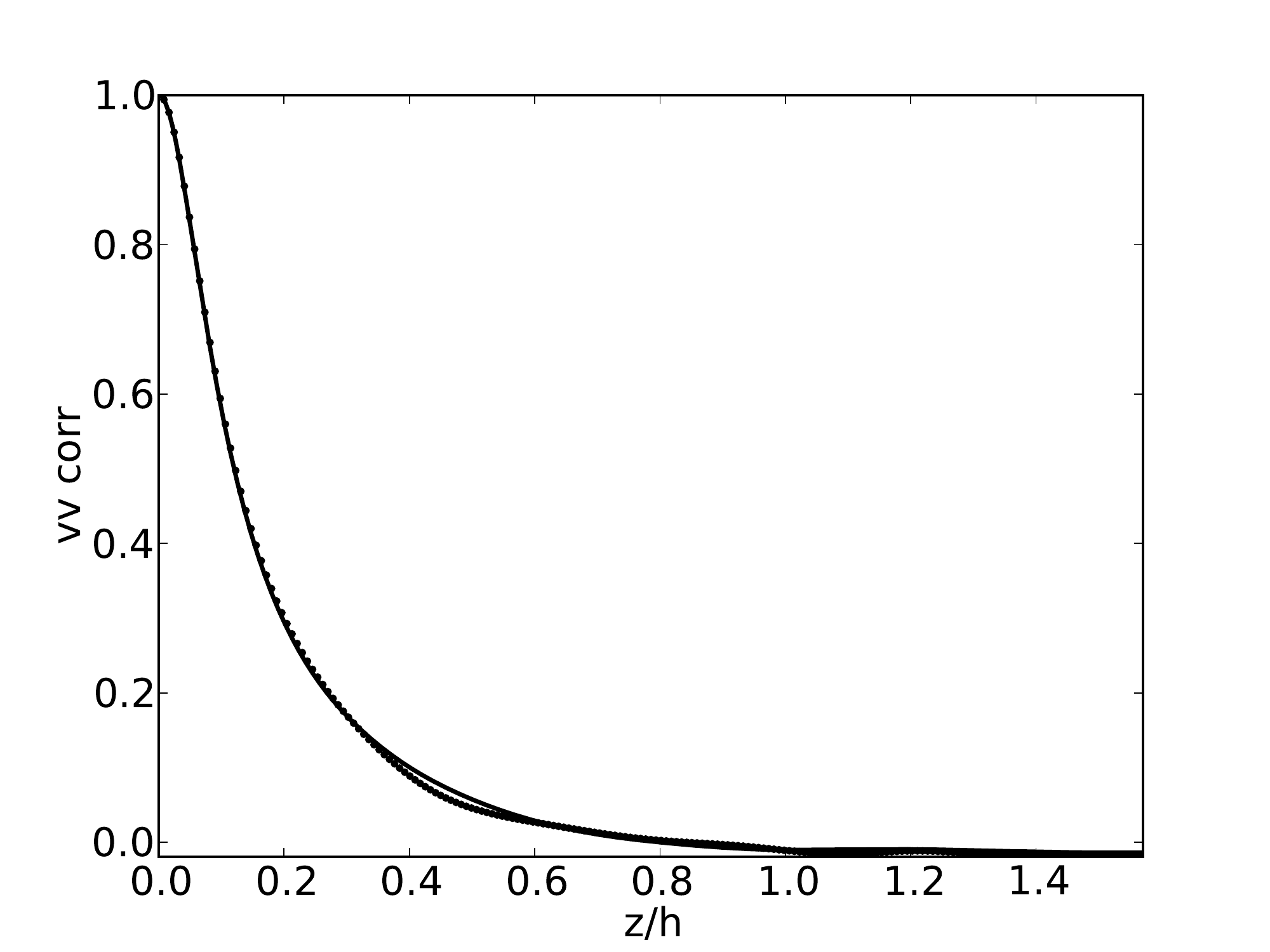}\includegraphics[scale=0.3]{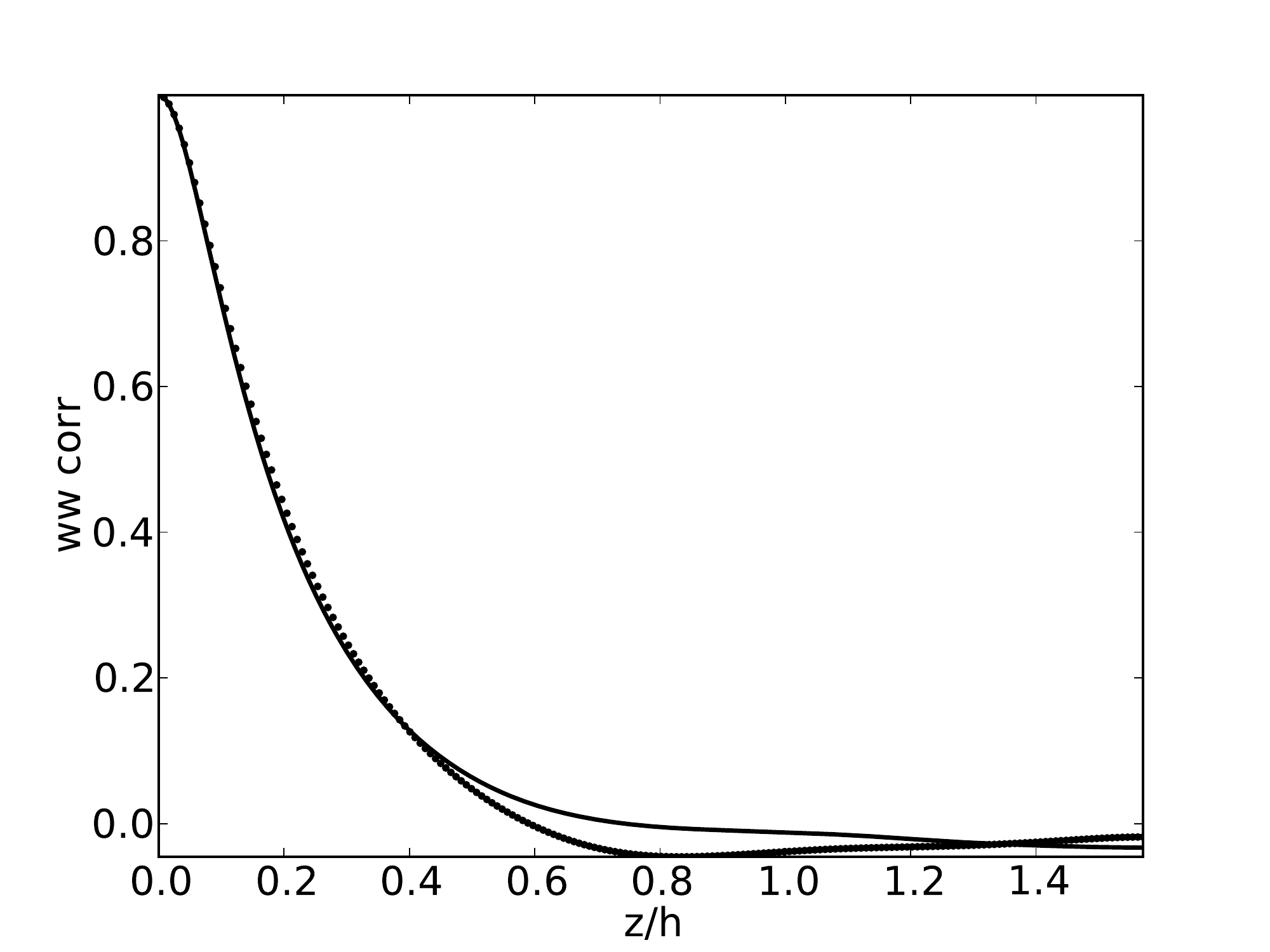}
\par\end{centering}

\caption{Further validation of run of Table \ref{tab:secn5.1-validation} against
data from \cite{MoserKimMansour1999}, which is dotted/thicker. \label{fig:secn5-validation-2}}
\end{figure}

The correlations shown in Figure \ref{fig:secn5-validation-2} are
another validation check. The correlations converge quite slowly,
with errors inversely proportional to the time of integration or worse.
Here too the agreement is quite good. In general, MKM data seems to
have slightly stronger nearby correlations compared to that of Table
\ref{tab:secn5.1-validation}. The longer time of integration employed
here has eliminated the slight artifact from nearby correlations.

\begin{table}
\begin{centering}
\begin{tabular}{|c|c|c|c|c|c|c|c|c|}
\hline 
$Re$ & $Re_{\tau}$ & $\Lambda_{x}$ & $\Lambda_{z}$ & $L/M/N$ & $dx^{+}/dy_{max}^{+}/dz^{+}$ & $T_{init}u_{\tau}/h$ & $Tu_{\tau}/h$ & CFL\tabularnewline
\hline 
\hline 
$80,000$ & $2391$ & $4/2\pi$ & $2/2\pi$ & $1024/1024/1024$ & $9.3/7.3/4.7$ & $0.1$ & $2.78$ & $0.25$\tabularnewline
\hline 
$80,000$ & $2385$ & $4/2\pi$ & $2/2\pi$ & $1024/1024/1024$ & $9.3/7.3/4.6$ & $2.88$ & $2.06$ & $0.25$\tabularnewline
\hline 
\end{tabular}
\par\end{centering}

\caption{Run parameters for two runs to test elimination of transients. The
columns have the same meaning as in Table \ref{tab:secn5.1-validation}.
\label{tab:secn5.2-transients}}
\end{table}

\begin{figure}
\begin{centering}
\includegraphics[scale=0.3]{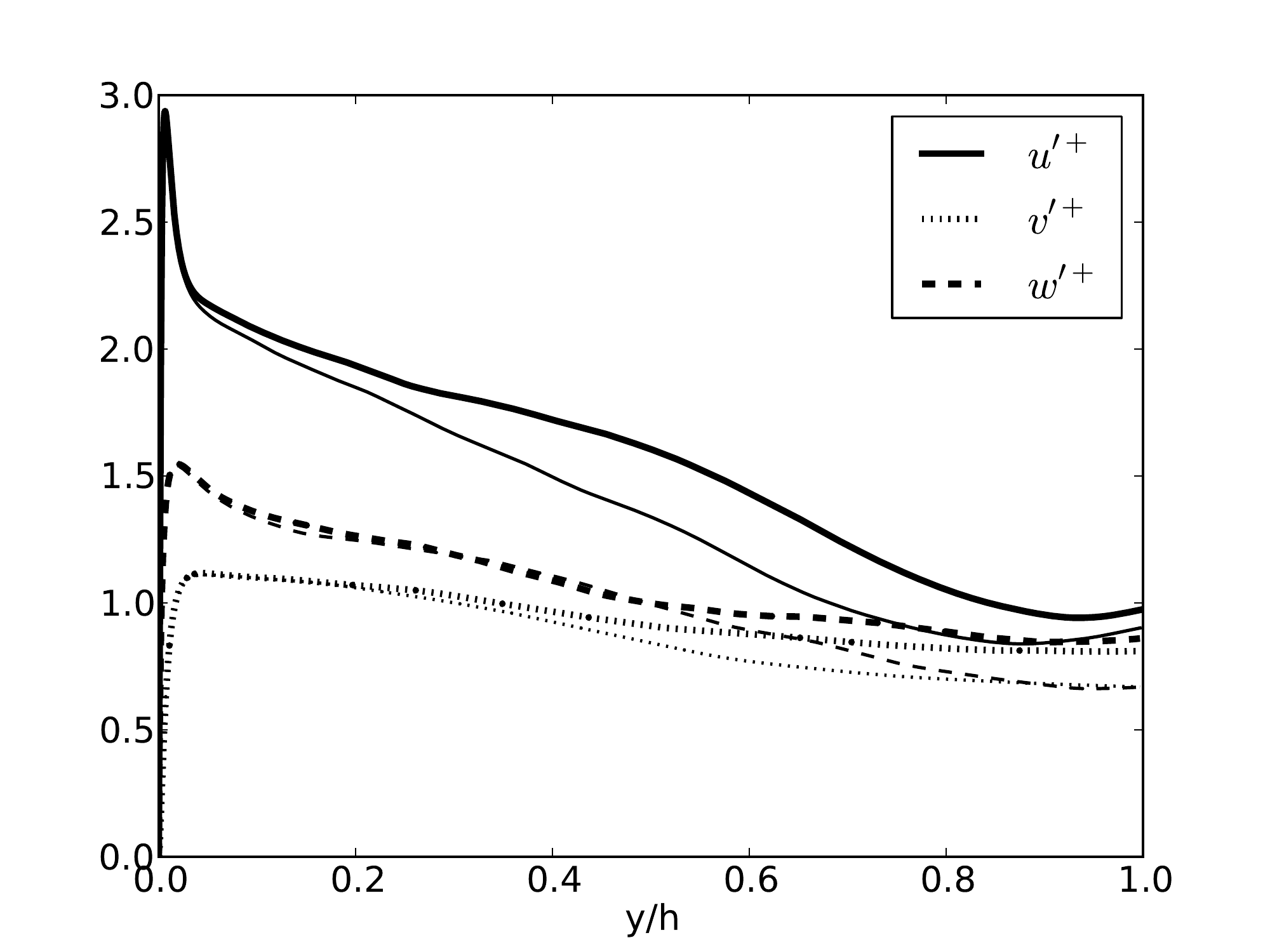}\includegraphics[scale=0.3]{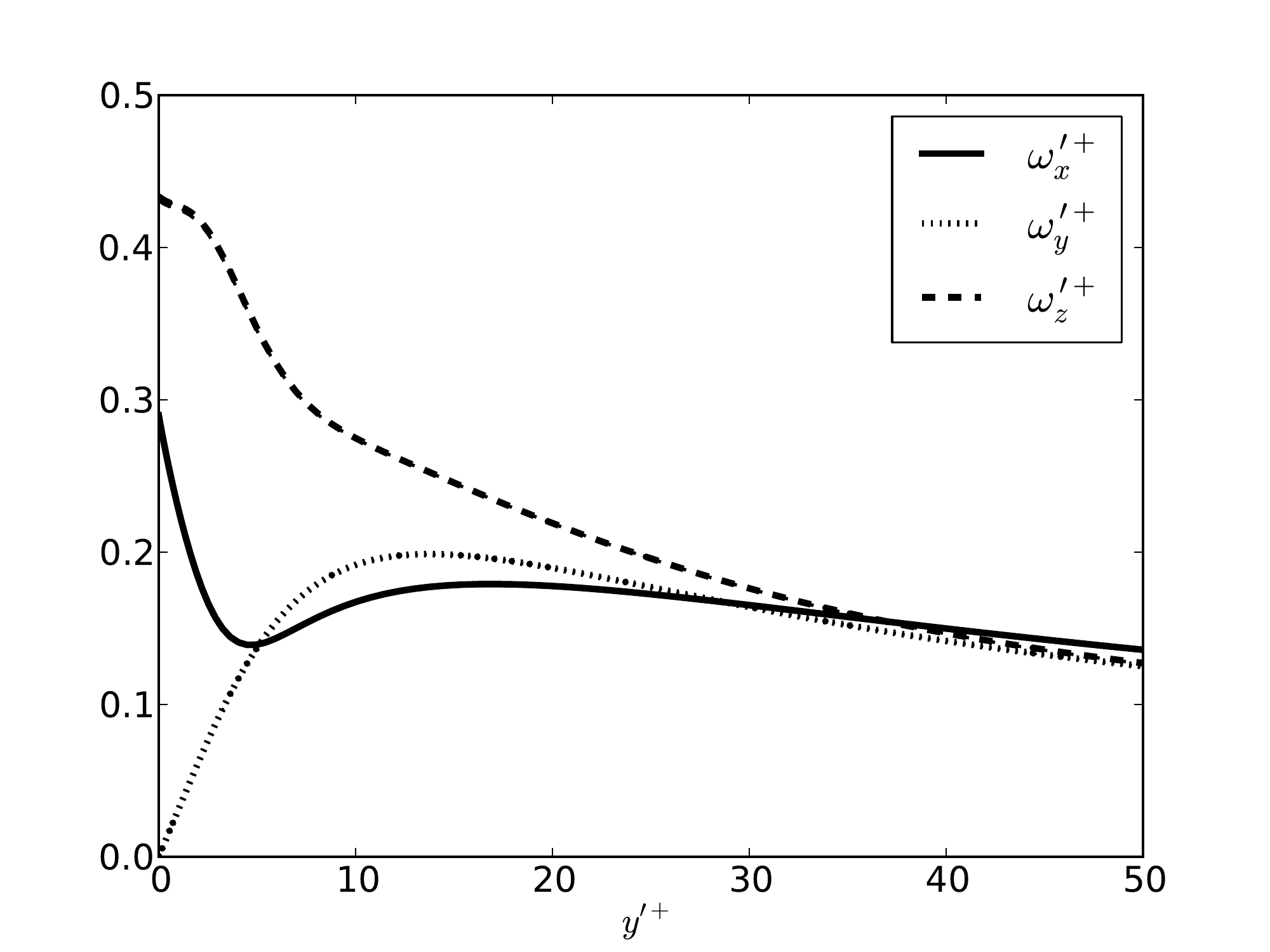}
\par\end{centering}

\begin{centering}
\includegraphics[scale=0.3]{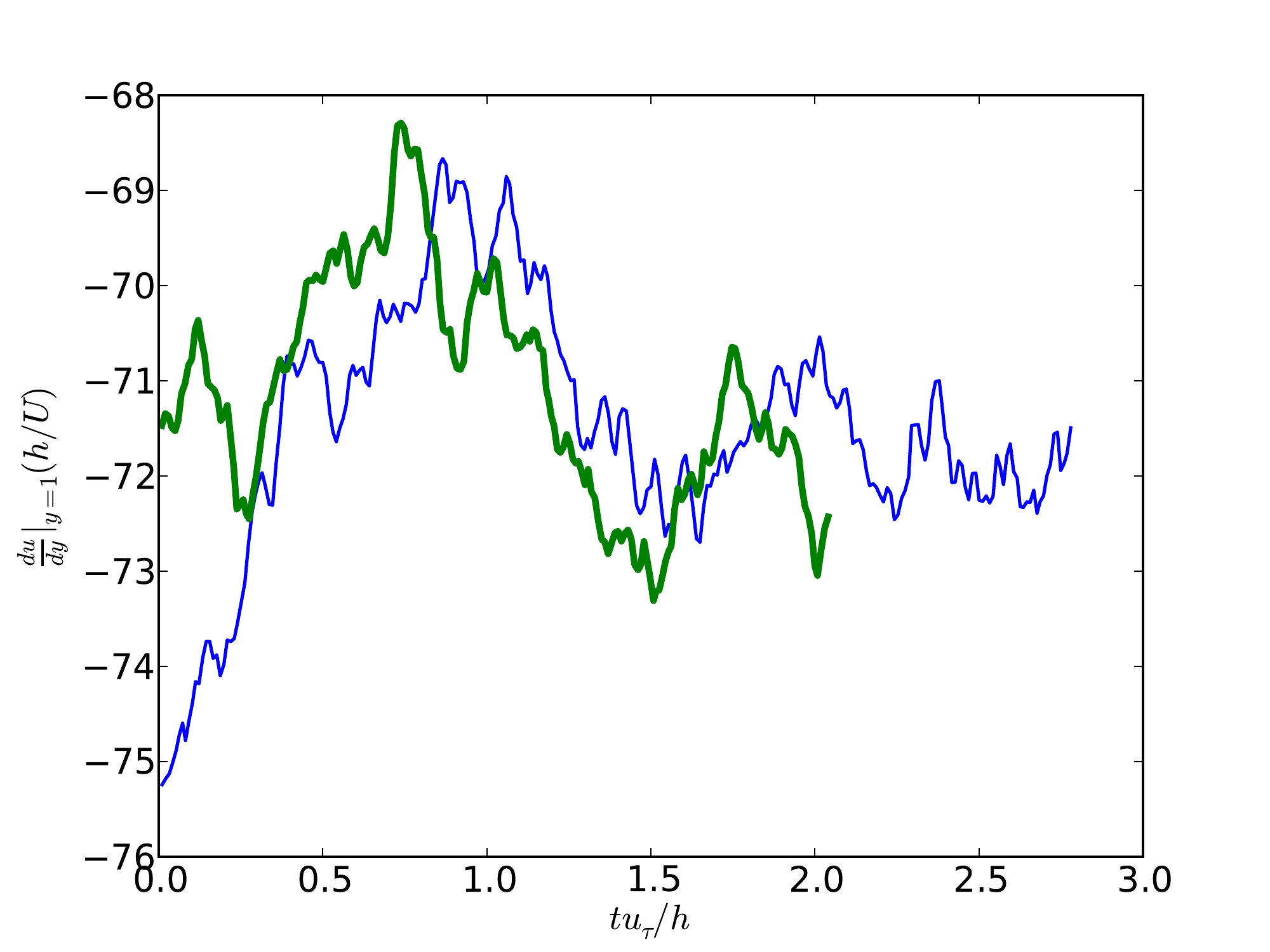}
\par\end{centering}

\caption{Comparison of the two runs of Table \ref{tab:secn5.2-transients}.
A longer run is used to eliminate transients in the second run and
plots corresponding to it are thicker.\label{fig:secn5.2-transients}}
\end{figure}

The purpose of the runs of Table \ref{tab:secn5.2-transients} is
to assess how short a $T_{init}$ for eliminating transients suffices
at the highest Reynolds number and the finest mesh. In those two runs,
it takes approximately $10^{5}$ time steps to integrate for a single
eddy turnover time $\tau_{eddy}$ and each step takes $16$ seconds.
Thus a reduction in $T_{init}$ is of much value.

Both runs began with an initial state that was generated at the coarser
$512^{3}$ resolution. In the first run, $T_{init}=0.1\tau_{eddy}$
and in the second run $T_{init}=2.88\tau_{eddy}$. The times $T$
used to gather statistics were $2.78\tau_{eddy}$ and $2.06\tau_{eddy}$,
respectively. Typically, $T\geq8\tau_{eddy}$ is needed for reliable
statistics \cite{HoyasJimenez2006,HoyasJimenez2008}, but here we
settle for smaller $T$ to better assess the effect of $T_{init}$. 

The first two plots of Figure \ref{fig:secn5.2-transients} show that
larger $T_{init}$ in the second run does not lead to statistics of
better quality. In fact, the turbulence intensities, especially $u^{\prime+}$,
are better converged for the first run most probably because it is
gathering statistics using a larger $T$.

The third plot of the same figure compares the shear at the wall as
a function of time for the two runs. In the first run, the magnitude
of the shear is greater than $75$ but begins to decay right away.
That is a telltale sign of the coarser grid origin of the initial
state. Coarser grid simulations are more turbulent and have greater
shear because there is less viscosity to smooth the flow in a coarser
simulation. But already at $t\approx0.5\tau_{eddy}$ the first run
seems to have reached a statistically steady turbulent state (or more
precisely, a sample from a statistically steady turbulent ensemble).
It appears that $T_{init}\in[0.5\tau_{eddy},1.0\tau_{eddy}]$ suffices
to eliminate transients from an initial state computed at lower resolution.

\begin{figure}
\begin{centering}
\includegraphics[scale=0.6]{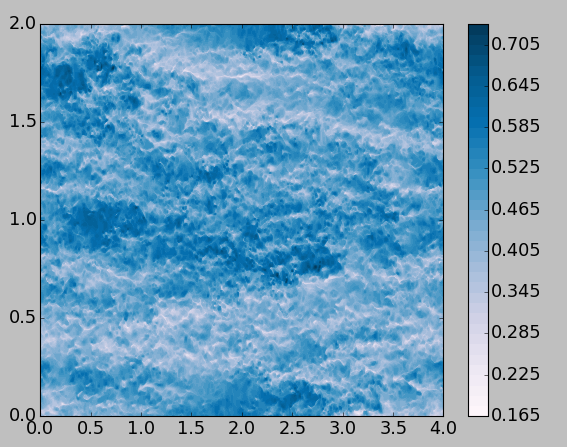}
\par\end{centering}

\caption{Contour plot of the $u$-component of the velocity at $y=0.97433$
with $y=1$ being the top wall of the channel at $t\approx6\tau_{eddy}$.}

\end{figure}

\section{Acknowledgments}

The author thanks Hans Johnston, Benson Muite, and Fabian Waleffe
for discussions and suggestions. This research was partially supported
by NSF grants DMS-1115277 and SCREMS-1026317.

\bibliographystyle{plain}
\bibliography{references}

\end{document}